\documentclass[floatfix,showpacs,superscriptaddress,prl,twocolumn]{revtex4}
\usepackage{amsbsy,amssymb,amsmath,bm}
\usepackage{graphicx,color}
\usepackage{amsfonts}
\usepackage{amssymb}
\usepackage{amsmath}

\input{tcilatex}

\begin{document}

\title{Lyapunov vs. Geometrical Stability Analysis of the Kepler and the
Restricted Three Body Problems}
\author{A. Yahalom }
\affiliation{\textit{Department of Electrical and Electronic Engineering, Ariel
University Center of Samaria, Ariel 40700, Israel}}
\author{J. Levitan}
\affiliation{\textit{Applied Physics Department, Ariel University Center of Samaria,
Ariel 40700, Israel}}
\email{levitan@ariel.ac.il}
\affiliation{\textit{Department of Physics, Technical University of Denmark, Lyngby 2800,
Denmark}}
\author{M. Lewkowicz}
\affiliation{\textit{Applied Physics Department, Ariel University Center of Samaria,
Ariel 40700, Israel}}
\author{L.Horwitz}
\affiliation{\textit{Applied Physics Department, Ariel University Center of Samaria,
Ariel 40700, Israel}}
\affiliation{\textit{School of Physics, Tel Aviv University, Tel Aviv 69978, Israel}}
\affiliation{\textit{Department of Physics, Bar-Ilan University, Ramat Gan 52900, Israel}}
\date{\today }

\begin{abstract}
In this letter we show that although the application of standard Lyapunov
analysis predicts that completely integrable Kepler motion is unstable, the
geometrical analysis of Horwitz et al [1] predicts the observed stability.
This seems to us to provide evidence for both the incompleteness of the
standard Lyapunov analysis and the strength of the geometrical analysis.
Moreover, we apply this approach to the three body problem in which the
third body is restricted to move on a circle of large radius which induces
an adiabatic time dependent potential on the second body. This causes the
second body to move in a very interesting and intricate but periodic
trajectory; however, the standard Lyapunov analysis, as well as methods
based on the parametric variation of curvature associated with the Jacobi
metric, incorrectly predict chaotic behavior. The geometric approach
predicts the correct stable motion in this case as well.
\end{abstract}

\maketitle

\subsection{I. Introduction}

Several relatively recent papers have used geometric approaches to describe
Hamiltonian chaos, implementing tools from Riemannian geometry \cite{PRE47,
PRE48, PRE51}. The natural motions of Hamiltonian systems are viewed as
geodesics on a Riemannian space associated with a metric $g_{ij}$ (often
associated with either the Jacobi or the Eisenhart metric). Stable motions
are thus defined by the curvature properties of the manifold.

Casetti \textit{et. al.} \cite{PRE48} derived from the Jacobi--Levi-Civita
(JLC) equation an effective stability equation which formally describes a
stochastic oscillator. They conjectured that an \textquotedblleft
average\textquotedblright\ global geometric property should provide
information about the \ degree of chaos and applied the geometric method to
the Fermi-Pasta-Ulam beta-model and to a chain of coupled oscillators.

Safaai and Saadat \cite{SafaaiSaadat} have used the geometric method
developed in \cite{PRE47, PRE48, PRE51} in an attempt to characterize the
behaviour of the restricted three body problem in celestial mechanics. They
conclude that the fluctuations of the curvature of the manifold along the
geodesics yield parametric instability of the trajectories and are
associated with chaos. They also calculated a positive Lyapunov exponent. We
point out here that our simulation demonstrates stability for these orbits.
We discuss the application of a recently developed geometrical criterion for
stability to this problem, and show that it correctly predicts stability.

It has recently been shown \cite{Horwitz} (to be called HBLSL) that there is
a possibility to characterize instability in Hamiltonian systems by a
geometrical approach which takes its point of origin in the curvature
associated with a conformal Riemann metric tensor essentially different from
the Jacobi metric, which is applicable to a large class of potential models
(see also \cite{Li}). This approach appears to be more sensitive than
computing Lyapunov exponents or the use of the Jacobi metric (see the book
of Pettini \cite{PettiniBook} for a review of important results obtained
with methods related to the use of the Jacobi metric). This enables one to
associate instability with a negative dynamical curvature appropriate for
the geodesic motion different from that defined by the Jacobi metric, for
which the line element is proportional to the time $t$ \ and not, as for the
Jacobi metric, the action. This establishes a natural connection between
chaotic Hamiltonian flows and Anosov flows \cite{Anosov} directly in the
time domain.

Additional problems in celestial mechanics can also be addressed with the
HBLSL method, such as planetary stability and the Pioneer anomaly (see for
example S. G. Turyshev and V.T. Toth \cite{Turyshev}).

\subsection{\textit{\ }II. The Geometrical Method of HBLSL}

In \cite{Horwitz} the stability of a Hamiltonian system of the form (we use
the summation convention)

\begin{equation}
H=\frac{\mathbf{p}^{2}}{2M}+V\left( x\right) =\frac{p^{i}p^{j}}{2M}\delta
_{ij}+V\left( x\right)  \label{1}
\end{equation}%
was studied, where $V$ is a function of space variables alone, by
introducing a second Hamiltonian of the type considered by Gutzwiller \cite%
{Jacobi, Gutzwiller}

\begin{equation}
H_{G}=\frac{1}{2M}g_{ij}p^{i}p^{j}  \label{2}
\end{equation}%
where $g_{ij}$ is a function of the coordinates alone. Hamilton's equations
applied to Eq. (2) result in the geodesic form

\begin{equation}
\ddot{x}_{l}=-\Gamma _{l}^{mn}\dot{x}_{m}\dot{x}_{n};  \label{3}
\end{equation}%
where%
\begin{equation}
\Gamma _{l}^{mn}=\frac{1}{2}g_{lk}\{\frac{\partial g^{km}}{\partial x_{n}}+%
\frac{\partial g^{kn}}{\partial x_{m}}-\frac{\partial g^{nm}}{\partial x_{k}}%
\},  \label{4}
\end{equation}%
and $g^{ij}$ is the inverse of $g_{ij}$ .

Horwitz, et al \cite{Horwitz} take $g_{ij}$ to be of conformal form%
\begin{equation}
g_{ij}=\frac{E}{E-V\left( x\right) }\delta _{ij}  \label{5}
\end{equation}%
on the hypersurface defined by $H=E=constant$, resulting in a formal
equivalence between the two Hamiltonians (\ref{1}) and (\ref{2}) on the same
energy shell $E$.

The velocity field, defined by%
\begin{equation}
\dot{x}^{j}\equiv g^{ji}\dot{x}_{i}=\frac{p^{j}}{M},  \label{6}
\end{equation}%
satisfies the geodesic equation%
\begin{equation}
\ddot{x}^{l}=-\mathcal{M}_{mn}^{l}\dot{x}^{m}\dot{x}^{n}  \label{7}
\end{equation}%
where%
\begin{equation}
\mathcal{M}_{mn}^{l}\equiv \frac{1}{2}g^{lk}\frac{\partial g_{nm}}{\partial
x^{k}}.  \label{8}
\end{equation}

The quantity (8) is a connection form, that is, it satisfies the
requirements for the construction of a covariant derivative, but it is not
compatible with the metric $g_{ij}$. It is therefore not a Christoffel
symbol, but it may be derived as well by parallel transport on the
Gutzwiller space and transformation to the coordinates $x^{j}$. The
coordinates $x^{j}$ correspond to the manifold for which the velocity field
is given by (6). This correspondence was discussed in \cite{Horwitz} and
will be discussed in more detail in \cite{HorwitzPrep}.

It is shown in \cite{Horwitz} that in the coordinates for which $g_{ij}$ has
the form (\ref{5}), Eq. (\ref{7}) coincides with the Hamilton equations
obtained from the Hamiltonian form (1), and thus provides a \textit{%
geometrical embedding} of this Hamiltonian motion. The curvature associated
with the connection form $\mathcal{M}$ is relevant to the dynamical
stability of the Hamiltonian motion rather than the curvature associated
with the connection form $\Gamma $ defined in (\ref{4}). The geodesic
deviation $\xi ^{l}$ between two such orbits $x^{\prime }(t)$ and $x(t)$
satisfies 
\begin{equation}
\frac{D_{\mathcal{M}}^{2}}{D_{\mathcal{M}}t^{2}}\xi ^{l}=R_{\mathcal{M}%
qmn}^{l}\dot{x}^{q}\dot{x}^{n}\xi ^{m},  \label{9}
\end{equation}%
where $\xi ^{l}=x^{\prime l}(t)-x^{l}(t).$

Here, the covariant derivative of $\xi ^{l}$ is defined by%
\begin{equation}
\xi _{;n}^{l}=\frac{\partial \xi ^{l}}{\partial x^{n}}+\mathcal{M}%
_{nm}^{l}\xi ^{m}  \label{10}
\end{equation}%
and$\frac{D_{\mathcal{M}}}{D_{\mathcal{M}}t}$is the corresponding covariant
derivative in$\ t$ along the instantaneously approximate common orbits,
i.e., 
\begin{equation}
\frac{D_{\mathcal{M}}}{D_{\mathcal{M}}t}\xi ^{l}=\dot{\xi}^{l}+\mathcal{M}%
_{nm}^{l}\dot{x}^{n}\xi ^{m}.  \label{11}
\end{equation}

What may be called the 'dynamical curvature' appearing in (9) is given by

\begin{equation}
R_{\mathcal{M}qmn}^{l}=\frac{\partial \mathcal{M}_{qm}^{l}}{\partial x^{n}}-%
\frac{\partial \mathcal{M}_{qn}^{l}}{\partial x^{m}}+\mathcal{M}_{qm}^{k}%
\mathcal{M}_{nk}^{l}-\mathcal{M}_{qn}^{k}\mathcal{M}_{mk}^{l}.  \label{12}
\end{equation}%
With the conformal metric in noncovariant form (5), the geodesic deviation
equation (9) becomes%
\begin{equation}
\frac{D_{\mathcal{M}}^{2}}{D_{\mathcal{M}}t^{2}}\xi =-\mathcal{V}P\xi ,
\label{13}
\end{equation}%
where the matrix $\mathcal{V}$ is given by%
\begin{equation}
\mathcal{V}_{li}=\frac{3}{M^{2}v^{2}}\frac{\partial V}{\partial x^{l}}\frac{%
\partial V}{\partial x^{i}}+\frac{1}{M}\frac{\partial ^{2}V}{\partial
x^{l}\partial x^{i}}  \label{14}
\end{equation}

and%
\begin{equation}
P^{ij}=\delta ^{ij}-\frac{v^{i}v^{j}}{v^{2}}  \label{15}
\end{equation}%
with $v^{i}=$ $\dot{x}^{i}$, defining a projection into a plane (in the
two-dimensional examples we shall treat here, just a line) orthogonal to $%
v^{i}$.

Computer investigations \cite{Ben Zion} of this local criterion show that
local modifications of the Hamiltonian in regions where negative eigenvalues
occur can be used to control the stability of the system. Removal or
modification of the nonlinear and symmety breaking terms in just those local
regions have dramatic effects on the Poincar\'{e} plots, completely
stabilizing the global motion in the examples studied there.

It was argued in \cite{Horwitz} that instability should occur if at least
one of the eigenvalues of $\mathcal{V}$ is negative. This criterion was
found to be effective for all of the examples we have studied \cite{Horwitz}%
. Our experience with these examples indicates that positive eigenvalues are
associated with stability. It is an interesting question, of course, to
study whether there could be parametric instability associated with these
positive eigenvalues (\textit{e.g.},\cite{PRE48}). We remark that the
investigations of \cite{SafaaiSaadat} indicated a parametric instability in
a system that was in fact stable, as our simulations show.

\subsection{III. Time Dependent Potentials}

The analysis given above was carried out with a time independent potential
function $V$. In order to analyze the stability of the restricted three body
problem using the methods discussed in Section II, we first study the effect
of a weak time dependence in $V$. In this case $E$ is not precisely
conserved, but we shall make an adiabatic approximation in which we consider 
$E$ to be time independent. The time dependence of $V$ appears in the form
of the geodesics generated by $H_{G}$ as an additional term with partial
time derivative of the metric, but in the geometrical imbedding of the
Hamiltonian motion, the geodesic evolution of the velocity vector field has
the same form (7) as for the time independent potential problem. The
formulation of geodesic deviation in terms of the second covariant
derivative, however, as we shall see below, introduces another time
derivative, introducing an additional term in the stability matrix.
Referring back to the discussion of the previous Section, Eq.(9) is obtained
by first computing $\ddot{\xi}^{l}$ with the help of (7). The resulting
equation is simplified and rendered more transparent in its physical
implication by re-expressing it in terms of the second covariant derivative.
Although Eq.(7) is unaffected in form in the presence of a time dependence
in $V$, the equation for the second covariant derivative, Eq.(9), now becomes

\begin{equation}
\frac{D_{\mathcal{M}}^{2}}{D_{\mathcal{M}}t^{2}}\xi ^{l}=R_{\mathcal{M}%
qrp}^{l}\dot{x}^{q}\dot{x}^{p}\xi ^{r}+\frac{\partial \mathcal{M}_{rp}^{l}}{%
\partial t}\dot{x}^{p}\xi ^{r},  \label{16}
\end{equation}%
which is just Eq.(9) with the additional term%
\begin{equation}
\frac{\partial \mathcal{M}_{rp}^{q}}{\partial t}\dot{x}^{p}\xi ^{r}=\frac{1}{%
2}\frac{\partial ^{2}g_{rp}}{\partial t\partial x_{q}}\dot{x}^{p}\xi ^{r}
\label{17}
\end{equation}%
Using the conformal form of $g_{ij}$ with time dependent potential, the
right hand side contains

\begin{equation*}
\frac{\partial ^{2}g_{rt}}{\partial x_{q}\partial t}=\ \ \ \ \ \ \ \ \ \ \ \
\ \ \ \ \ \ \ \ \ \ \ \ \ \ \ \ \ \ \ \ \ \ \ \ \ \ \ \ \ \ \ \ \ \ \ \ \ \
\ \ \ \ \ \ \ \ \ \ \ \ \ \ \ \ \ \ \ \ \ \ 
\end{equation*}
\ 

\begin{equation}
\left[ \frac{2E}{\left( E-V\left( x,t\right) \right) ^{3}}\frac{\partial V}{%
\partial t}\frac{\partial V}{\partial x_{q}}+\frac{E}{\left( E-V\left(
x,t\right) \right) ^{2}}\frac{\partial ^{2}V}{\partial x_{q}\partial t}%
\right] \delta _{rt}  \label{18}
\end{equation}

Near the boundary of the physical region, these terms can become large, but
they are generally small away from these boundaries since they involve the
time derivative of $V$, in the same approximation in which $E$ is constant
(the consistency of this approximation will be discussed elsewhere). In our
computations here these conditions hold, and we therefore compute stability
using the same formulas as for the time independent problem. The modified
Eq.(16) will be discussed in greater detail in a succeeding publication.

\subsection{IV. Application to Restricted Three Body Problem and Kepler Limit%
}

We now consider the following Lagrangian of a restricted three body system: 
\begin{eqnarray}
L &=&T-V,\text{ \ \ \ }T=\frac{1}{2}m_{e}\left( \frac{d\vec{r}}{dt}\right)
^{2}  \notag \\
V &=&\frac{1}{2}m_{j}r_{j}^{2}\omega _{j}^{2}-\frac{4\pi ^{2}m_{e}}{r}-\frac{%
4\pi ^{2}m_{j}}{r_{j}}-\frac{4\pi ^{2}m_{e}m_{j}}{r_{ej}}.  \label{19}
\end{eqnarray}%
In this system three bodies are involved: a large mass $M_{S}$ (the "Sun")
at the origin (see Fig.\ref{3bp1}), a mass $m_{e}$ ("Earth") located at $%
\vec{r}$, for which we study the trajectory, and a third mass $m_{j}$ (a
"Jupiter") located at $\vec{r}_{j}$ circling the Sun, 
\begin{equation}
\vec{r}_{j}=r_{j}(cos(\omega _{j}t),sin(\omega _{j}t)),  \label{20}
\end{equation}%
with $\omega _{j}$ and $r_{j}$ assumed to be constant (the first and third
terms in $V$ are thus constant.). $r_{ej}$ denotes the distance between
"Earth" and "Jupiter". The time scale is in years, distances are in
astronomical units and masses are scaled by the Sun mass. In these natural
units $GM_{S}=4\pi ^{2}.$

\begin{figure}
  \begin{center}
    \includegraphics[width=\columnwidth]{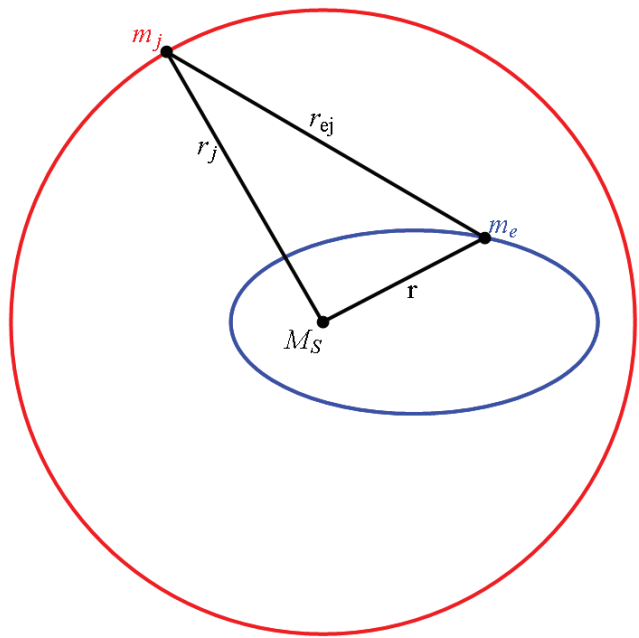}
	\caption{Depiction of the restricted
three body problem.}
  \end{center}
\end{figure}

We study the stability of the 'Earth' trajectory using the standard Lyapunov
analysis and also by the HBLSL geometrical approach. As shown in the
Appendix the Lyapunov exponents depend on the eigenvalues of the matrix $%
\mathcal{V}^{HS}$ (which we shall call the Hamiltonian stability matrix) 
\begin{equation}
\mathcal{V}_{kl}^{HS}=\frac{1}{m_{e}}\frac{\partial ^{2}V}{\partial
x_{k}\partial x_{l}},\text{ \ }k,l\in \{1,2\},\text{ \ }x_{1}\equiv x,\quad
x_{2}\equiv y.  \label{21}
\end{equation}

The standard Lyapunov eigenvalues $\lambda $ for the Hamiltonian motion
(obtained, as usual, by reducing the set of the second order Newtonian
equations to two sets of first order equations) are given by the square root
of minus the eigenvalues of $\mathcal{V}^{HS}$ (which we denote by $\lambda
_{HS}$), $\lambda =\pm \sqrt{-\lambda _{HS}}.$ Therefore, a positive
eigenvalue for $\mathcal{V}^{HS}$ implies Lyapunov stability, and a negative
eigenvalue implies Lyapunov instability (the resulting signs are therefore
the opposite of the usual Lyapunov convention).

In the HBLSL geometrical approach we determine the stability of the
trajectory by the eigenvalues of $\mathcal{V}$ in Eq.(14) which we rewrite
as: 
\begin{equation}
\mathcal{V}_{kl}=\frac{1}{m_{e}}\left[ \frac{3}{2(E-V)}\frac{\partial V}{%
\partial x_{k}}\frac{\partial V}{\partial x_{l}}+\frac{\partial ^{2}V}{%
\partial x_{k}\partial x_{l}}\right] .  \label{22}
\end{equation}
$E$ is now assumed to be adiabatically constant. By Eq.(13) a negative
eigenvalue indicates an unstable trajectory while a positive eigenvalue
indicates stability.

We study several cases distinguished in their 'Jupiter' mass (the Kepler
problem with zero Jupiter mass, the restricted Sun-Earth-Jupiter system with
a 'regular' Jupiter mass and a 'large Jupiter') and in their 'Earth' initial
conditions (the latter leading to different eccentricities in the Kepler
case). The planar motion that we analyze in these examples corresponds to
the known planetary configurations; this enables us to reduce the problem to
one with two coordinate degrees of freedom.

\subsubsection{The Kepler problem}

Letting the Jupiter mass equal to zero, $m_{j}=0,$ we treat the classical
Kepler problem. The Earth mass is taken as $m_{e}=2.96\cdot 10^{-6}.$ Three
cases with different initial conditions are analyzed.

\paragraph{Circular orbit}

A circular orbit is achieved by the initial conditions 
\begin{equation}
\vec{r}(0)=(1,0),\quad \frac{d\vec{r}}{dt}|_{t=0}=(0,2\pi ).
\label{initialcondcirc}
\end{equation}

The eigenvalues of the Hamilton stability matrix $\mathcal{V}^{HS}$ are $%
-8\pi ^{2}$ and $4\pi ^{2},$ whereas the two eigenvalues of the HBLSL
stability matrix $\mathcal{V}$ are both $4\pi ^{2}$. Thus according to
Lyapunov analysis the Kepler problem is unstable and could become chaotic.
This is clearly a false result since the Kepler problem is integrable. (Of
course, if one uses action-angle variables for the Kepler problem, it is
predicted to be stable; we have used the Cartesian coordinates in order to
illustrate that the Lyapunov method when applied without a careful choice of
coordinates, not always available, may lead to an incorrect prediction \cite%
{Berry}). The geometrical HBLSL approach predicts stability of the orbit.

\paragraph{Realistic eccentricity}

A more realistic orbit is obtained with the initial conditions 
\begin{equation}
\vec{r}(0)=(1-e,0),\frac{d\vec{r}}{dt}|_{t=0}=(0,2\pi \sqrt{\frac{1+e}{1-e}}%
),e=0.0167,  \label{initialcondreal}
\end{equation}%
where $e$ is the eccentricity of the Earth trajectory. One of the
eigenvalues $\lambda _{HS}$ is again negative, oscillating \ around its
'cicular' value $-8\pi ^{2}$, see Fig. 2, whereas the geometrical
eigenvalues are both positive, showing oscillating deviations from their
'circular' value $4\pi ^{2}$ due to the small changes in the orbital radius.%
\footnote[1]{%
Here and below we show always one relevant eigenvalue in the figures; the
other eigenvalues have a positive sign everywhere.} One of the eigenvalues
is shown in Fig. 3, with the negative deviations, whereas the second
eigenvalue bears positive deviations.

\begin{figure}
  \begin{center}
    \includegraphics[width=\columnwidth]{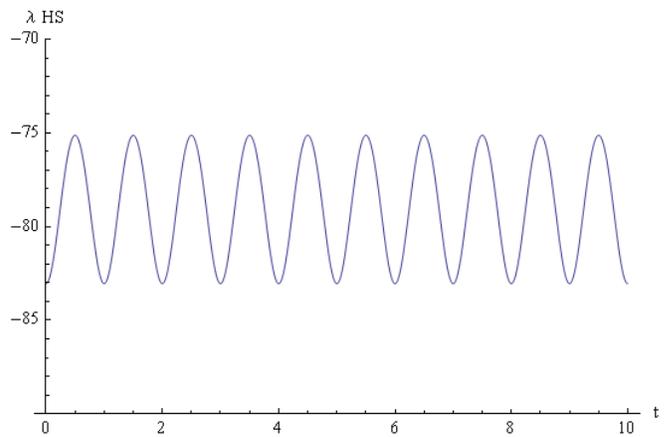}
	\caption{The eigenvalue of the Hamilton
stability matrix Eq.(21) as a function of time for Kepler problem, $m_{j}=0,$
with initial conditions (\protect\ref{initialcondreal}). Note that for the
circular orbit the HS eigenvalue is $-8\protect\pi ^{2}.$}
  \end{center}
\end{figure}

\begin{figure}
  \begin{center}
    \includegraphics[width=\columnwidth]{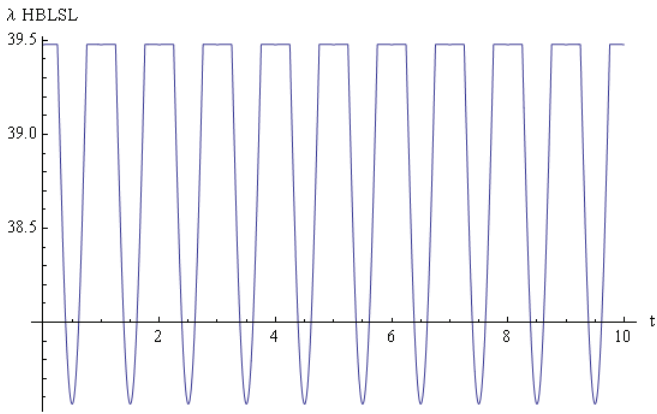}
	\caption{The eigenvalue of the HBLSL
stability matrix Eq.(22) as a function of time for the Kepler problem, $%
m_{j}=0,$ with initial conditions (\protect\ref{initialcondreal}). Note that
for the circular orbit the HBLSL eigenvalue is $4\protect\pi ^{2}.$}
  \end{center}
\end{figure}

\paragraph{Comparison with Ref.\protect\cite{SafaaiSaadat}}

In order to compare our results with those of \cite{SafaaiSaadat} we used
the same initial conditions: 
\begin{equation}
\vec{r}(0)=(1,0),\frac{d\vec{r}}{dt}|_{t=0}=(0,1).  \label{initialcondiran}
\end{equation}%
The small initial velocity causes a large 'eccentricity' in the ellipse-like
trajectory, which allows us to elucidate the question of stability. We
remark that the initial conditions that we have chosen, Eq.(\ref%
{initialcondreal}), take into account explicitly the eccentricity of the
elliptic orbit. The computations of \cite{SafaaiSaadat} used a shifted $x$
coordinate and a $y$ coordinate scaled by $2\pi $. The trajectory of the
'Earth' is depicted in Fig.4.

\begin{figure}
  \begin{center}
    \includegraphics[width=\columnwidth]{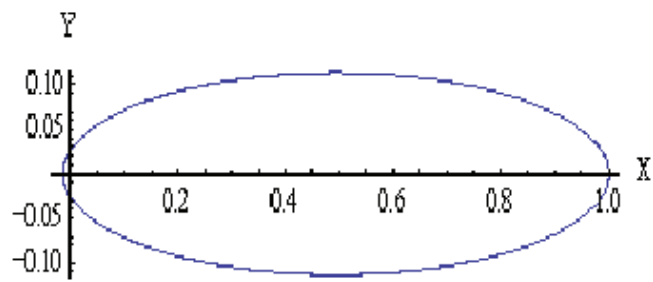}
	\caption{The Kepler orbit ($m_{j}=0)$
with initial conditions (\protect\ref{initialcondiran}).}
  \end{center}
\end{figure}

Fig. 5 reveals that one of the Hamiltonian stability eigenvalues is indeed
negative along the entire orbit although its negative value is less
pronounced between perihelia than at the perihelia.

\begin{figure}
  \begin{center}
    \includegraphics[width=\columnwidth]{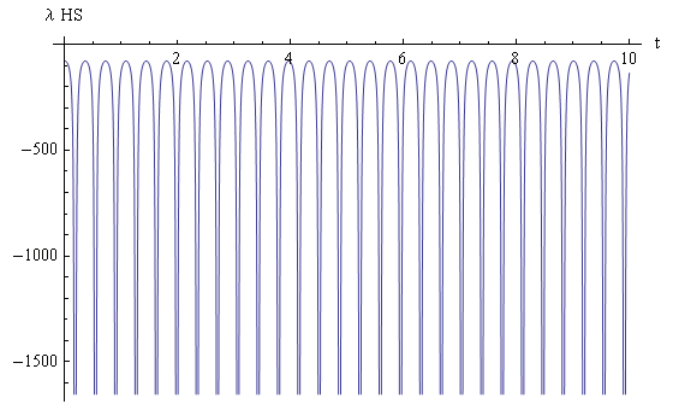}
	\caption{The eigenvalue of the Hamilton
stability matrix Eq.(21) as a function of time for the Kepler problem with
initial conditions (\protect\ref{initialcondiran})}
  \end{center}
\end{figure}

Yet the geometric HBLSL eigenvalue, depicted in Fig. 6 as a function of
time, is found to be positive in between perihelia, hence the trajectory is
predicted to be globally stable. The trajectory only spends a short time at
the unstable perihelia; this instability appears not to have an effect on
the behaviour of the orbit globally in this case. We infer that if such
local regions of negative eigenvalue are traversed in sufficiently short
time, or have sufficiently small extension, global instabilities will not be
observed (we are currently investigating this effect). This is a clear
indication that the geometrical HBLSL analysis is preferable.

\begin{figure}
  \begin{center}
    \includegraphics[width=\columnwidth]{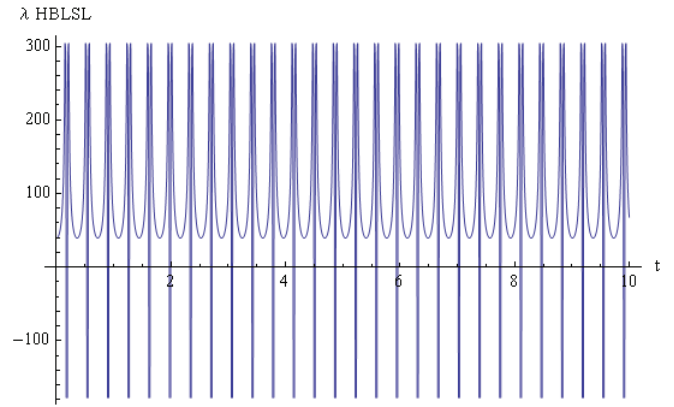}
	\caption{The eigenvalue of the HBLSL
stability matrix Eq.(22) for the restricted three body problem with $%
m_{j}=m_{j0}$ with initial conditions (\protect\ref{initialcondiran}).}
  \end{center}
\end{figure}

\subsubsection{The restricted three body problem}

\paragraph{The actual Jupiter mass}

In the second case we study we use the 'regular' Jupiter mass $m_{j0}=317.9$ 
$m_{e}=9.41\cdot 10^{-4},$ the mean Jupiter-Sun distance $r_{j}=5.203,$ and
the Jupiter orbital period of $11.86$ years.

Applying the initial conditions Eq.(\ref{initialcondreal}) that lead to the
realistic eccentricity in the Kepler case) we find that the Earth orbit is
not altered by the 'added' Jupiter. This is also true for the eigenvalues
depitcted in Figs.7 and 8. A minute oscillation of a 24 year period due to
Jupiter is apparent.

\begin{figure}
  \begin{center}
    \includegraphics[width=\columnwidth]{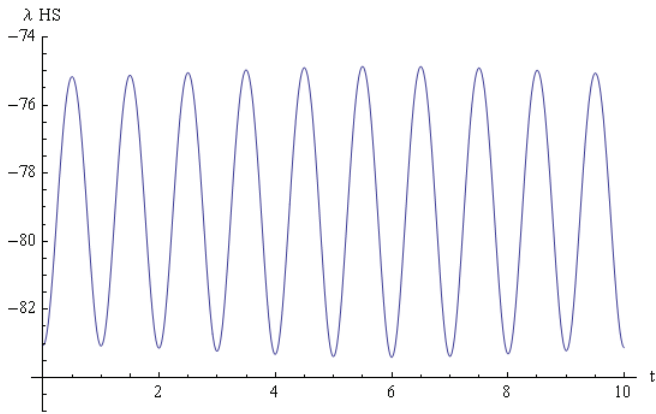}
	\caption{The eigenvalue of the Hamilton
stability matrix Eq.(21) for the restricted three body problem with $%
m_{j}=m_{j0}$ with initial conditions (\protect\ref{initialcondreal}).}
  \end{center}
\end{figure}

\begin{figure}
  \begin{center}
    \includegraphics[width=\columnwidth]{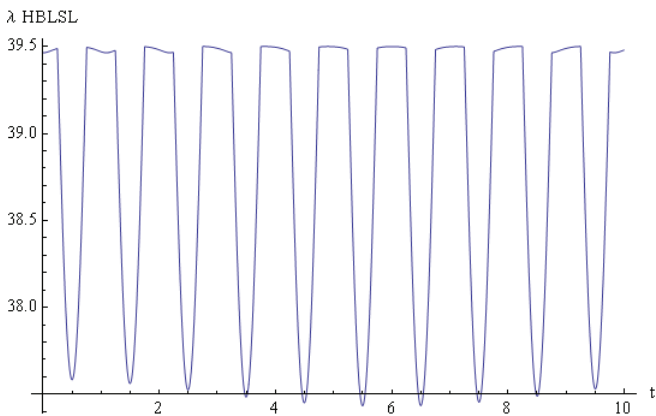}
	\caption{The eigenvalue of the HBLSL
stability matrix Eq.(22) for the restricted three body problem with $%
m_{j}=m_{j0}$ with initial conditions (\protect\ref{initialcondreal}).}
  \end{center}
\end{figure}

While employing the initial conditions of \cite{SafaaiSaadat} the 'added'
Jupiter does also not visibly modify the orbit and the eigenvalues as
compared with the relevant Kepler case.

\paragraph{A large Jupiter}

The third case deals with a 'large Jupiter', $m_{j}=1000m_{j0}.$ (We remark
that this value is close to the mass of the Sun, and therefore the dynamics
is close to that of a binary solar system. We shall discuss this point
further elsewhere.)

Again applying the initial conditions $\vec{r}(0)=(1-e,0),$ $\frac{d\vec{r}}{%
dt}|_{t=0}=(0,2\pi \sqrt{\frac{1+e}{1-e}})$ results in a knotty, yet
periodic trajectory, shown in Fig 9. The Hamilton stability eigenvalue is
again always negative along the orbit, depicted in Fig. 10, while the HBLSL
geometric eigenvalue along the orbits, Fig. 11, is mainly positive with
short negative excursions, similar to the Kepler case discussed above.

\begin{figure}
  \begin{center}
    \includegraphics[width=\columnwidth]{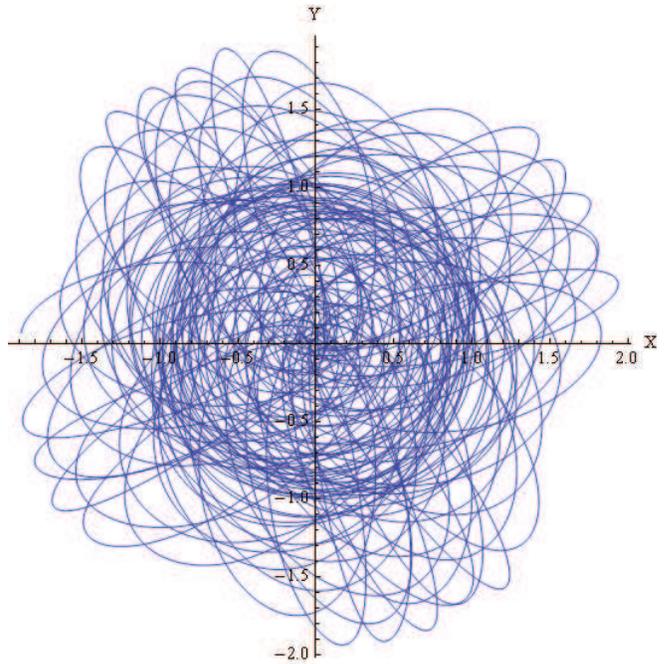}
	\caption{The earth orbit for the
restricted three body problem with $m_{j}=1000m_{j0}\ $during 100 time units
with the initial conditions Eq.(\protect\ref{initialcondreal}).}
  \end{center}
\end{figure}

\begin{figure}
  \begin{center}
    \includegraphics[width=\columnwidth]{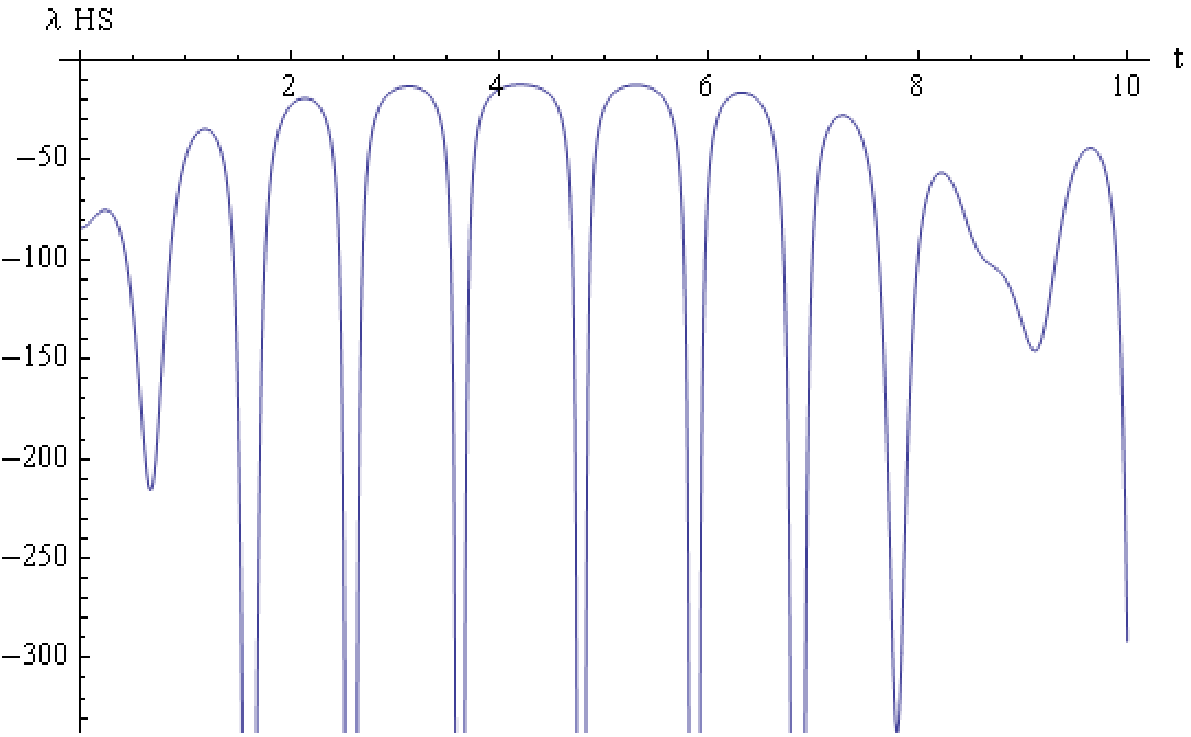}
	\caption{The eigenvalue of the Hamilton
stability matrix Eq.(21) as a function of time for the restricted three body
problem with $m_{j}=1000m_{j0}$ and initial conditions (\protect\ref%
{initialcondreal}).}
  \end{center}
\end{figure}

\begin{figure}
  \begin{center}
    \includegraphics[width=\columnwidth]{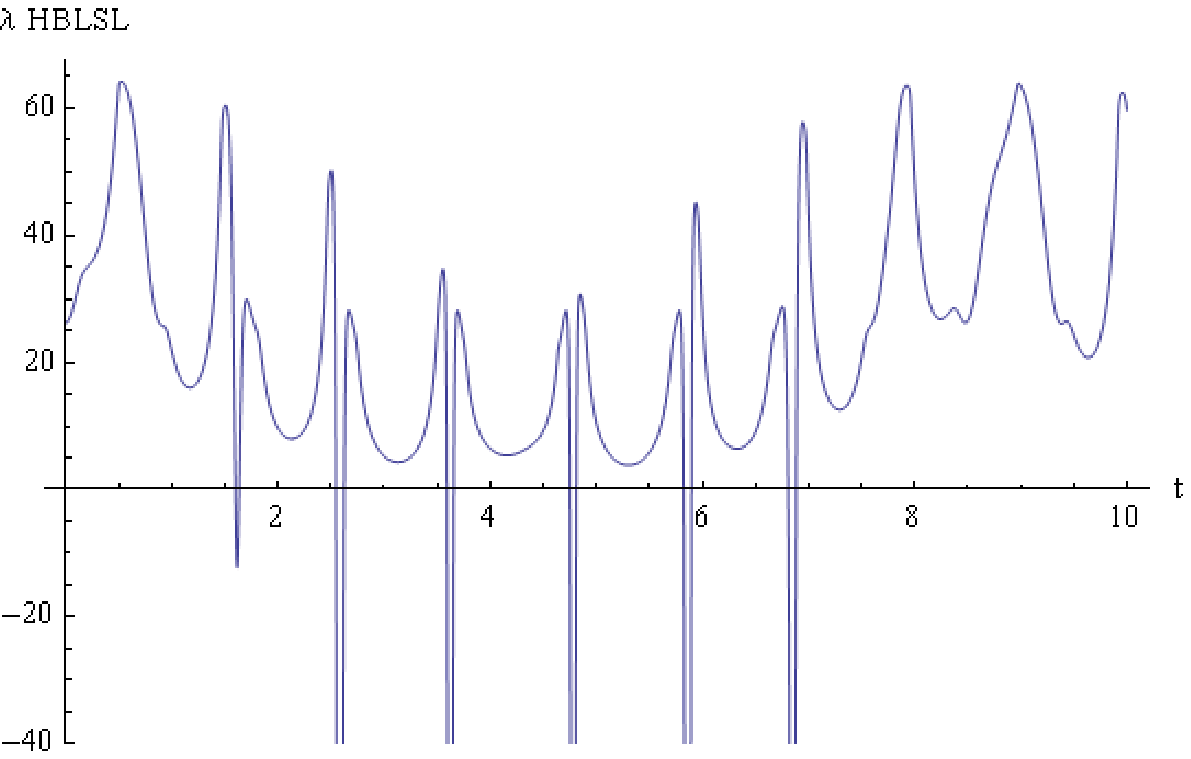}
	\caption{The eigenvalue of the HBLSL
stability matrix Eq.(22) as a function of time for the restricted three body
problem with $m_{j}=1000m_{j0}$ with initial conditions (\protect\ref%
{initialcondreal}).}
  \end{center}
\end{figure}

The initial conditions of \cite{SafaaiSaadat} with the 'large Jupiter'
create a complicated, but clearly periodic 'Earth' trajectory, Fig. 12. The
eigenvalues are given in Figs. 13 and 14. The same conclusions are obtained:
Lyapunov analysis predicts instability and possible chaos, while the
geometrical HBLSL analysis predicts stability.

\begin{figure}
  \begin{center}
    \includegraphics[width=\columnwidth]{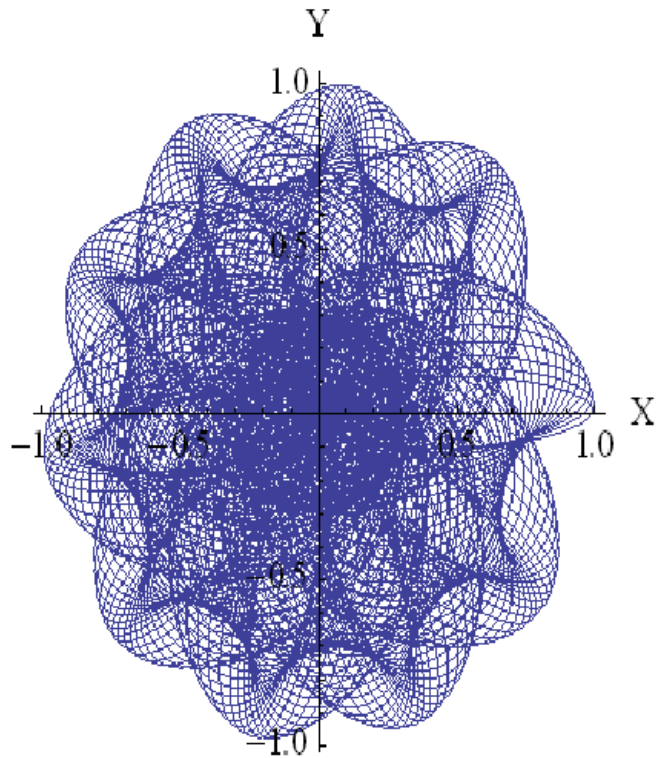}
	\caption{The earth orbit for the
restricted three body problem with $m_{j}=1000m_{j0}\ $during 100 time units
with initial conditions (\protect\ref{initialcondiran}).}
  \end{center}
\end{figure}

\bigskip

\begin{figure}
  \begin{center}
    \includegraphics[width=\columnwidth]{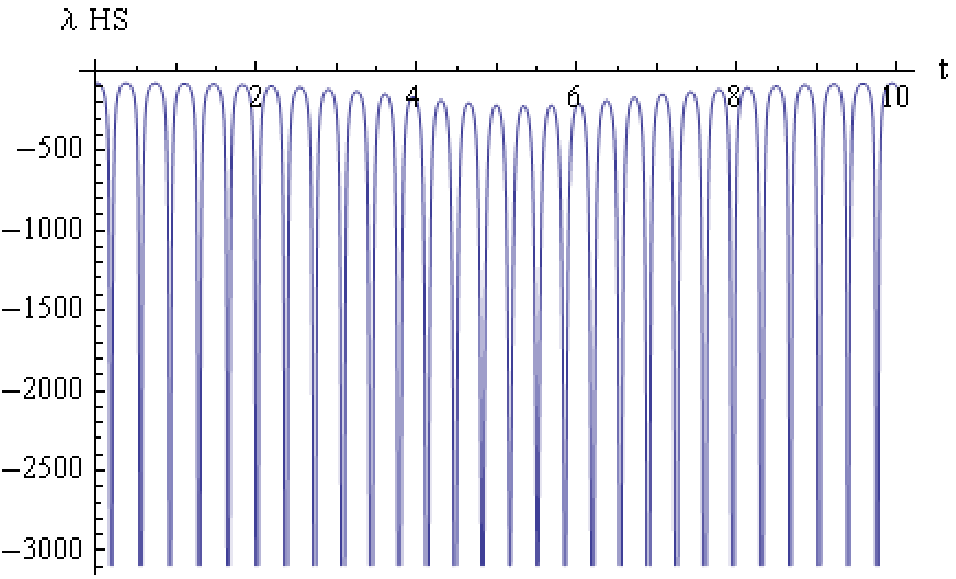}
	\caption{The eigenvalue of the Hamilton
stability matrix Eq.(21) as a function of time for the restricted three body
problem with $m_{j}=1000m_{j0}$ with initial conditions (\protect\ref%
{initialcondiran}).}
  \end{center}
\end{figure}

\begin{figure}
  \begin{center}
    \includegraphics[width=\columnwidth]{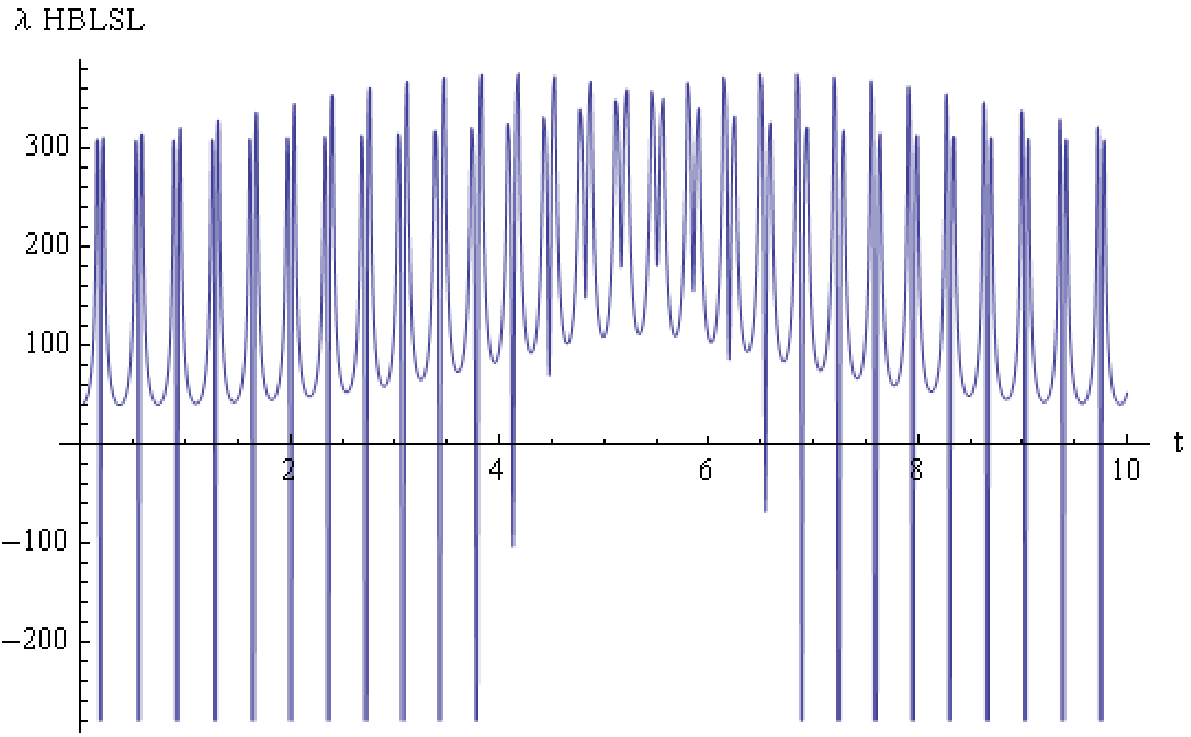}
	\caption{The eigenvalue of the HBLSL
stability matrix Eq.(22) as a function of time for the restricted three body
problem with $m_{j}=1000m_{j0}$ with initial conditions (\protect\ref%
{initialcondiran}).}
  \end{center}
\end{figure}

The application by Safaai \textit{et. al.} \cite{SafaaiSaadat} of the
methods of Casetti and Pettini using the criterion of stability under
parametric oscillations (see ref.\cite{PettiniBook}) also appear to predict
chaotic behaviour, which is not in agreement with our simulations.

\subsection{V. Conclusions}

The Lyapunov method for testing stability of Hamiltonian motion, as well as
the parametric oscillator extensions of the Jacobi metric criteria \cite%
{PRE47,PRE48,PRE51,PettiniBook} are not effective for the case of the
restricted three body problem. The Lyapunov procedure, involving a
linearization of the equations of motion in the neighborhood of a point on
the orbit, does not take into account the curvature of the surface on which
the system evolves, and we conjecture that this is the reason that the
Lyapunov criterion is ineffective in certain applications. Moreover, the use
of the Jacobi metric to define the geometry may also be weakened in its
effectiveness, since the invariant interval is the action rather than the
actual time; as pointed out in ref.\cite{PRE48}, the transformation changing
the parameter along the orbit from action to time has the effect of
essentially removing the geometrical embedding. This indicates, as for the
Lyapunov computation, that the curvature of the surface on which the motion
evolves may not be, in certain cases, sufficiently taken into account. The
HBSL method provides an embedding of the Hamiltonian motion in a geometrical
framework that appears to properly take into account the curvature of the
surface on which the motion evolves in the actual laboratory time.

The HBSL method correctly predicts the stability of the restricted three
body system. Moreover, the application of the Lyapunov method to the case $%
m_{j}\rightarrow 0$, the two body Kepler problem, predicts instability,
whereas the HBLSL criterion is consistent with stability. We remark that
there are clearly cases where adiabatic perturbation can result in strong
chaos \cite{Elskens}. In our analysis, for Hamiltonian systems, even if the
Hamiltonian function is constant to a good approximation, the contribution
given by (18) could have a large effect if the system approaches physical
boundaries, such as could happen in the problem, for example, of the Duffing
oscillator which has a hyperbolic fixed point. This does not occur in the
case we are studying here. The effects of (18) when it is not negligible are
under study.

We further remark that the form (19) can be easily generalized to treat the
effect of several (unperturbed) planets on the motion of the two body
subsystem, and the method can therefore be extended to a wider class of
problems in celestial mechanics (as well as satellite motion).

We conclude, with this application and others previously treated \cite%
{Horwitz,Li} that the recently discovered HBLSL criterion is a powerful and
useful tool for the analysis of the stability of Hamiltonian systems.

Acknowledgements. We thank the referees for a careful reading of the
manuscript and their helpful comments. We are indebted to M. Schiffer, E.
Kalderon and S. Shnider for valuable discussions.

\section{Appendix}

The Newtonian equation of motions are

\begin{equation}
\ddot{x}_{i}=-\frac{\partial V\left( x\right) }{\partial x_{i}}  \tag{A1}
\end{equation}

Substituting $y_{i}=\dot{x}_{i}$ results in a system of two first-order
equations for each degree of freedom $i$

\begin{eqnarray}
\dot{x}_{i} &=&y_{i}  \notag \\
\dot{y}_{i} &=&-\frac{\partial V\left( x\right) }{\partial x_{i}} 
\TCItag{A2}
\end{eqnarray}

Consider a new set of variables which are functions of the $x_{i}$ and $%
y_{i} $, 
\begin{equation*}
q_{k}=q_{k}\left( x_{i},y_{i}\right) .
\end{equation*}

Expanding $q_{k}$ at $r_{j}+\delta r_{j}$, where the $r_{j}$ denote $%
x_{i},y_{i},$%
\begin{equation*}
q_{k}+\delta q_{k}=q_{k}\left( r_{j}+\delta r_{j}\right) =q_{k}\left(
r_{j}\right) +\frac{\partial q_{k}}{\partial r_{j}}\delta r_{j}
\end{equation*}

and differentiating in time: $\dot{q}_{k}+\delta \dot{q}_{k}=\dot{q}%
_{k}\left( r_{j}+\delta r_{j}\right) =\dot{q}_{k}\left( r_{j}\right) +\frac{%
\partial \dot{q}_{k}}{\partial r_{j}}\delta r_{j}$ gives

\begin{equation*}
\delta \dot{q}_{k}=\frac{\partial \dot{q}_{k}}{\partial r_{j}}\delta r_{j}
\end{equation*}

Explicitly

\begin{equation}
\left( 
\begin{array}{c}
\delta \dot{x}_{1} \\ 
\delta \dot{x}_{2} \\ 
\delta \dot{y}_{1} \\ 
\delta \dot{y}_{2}%
\end{array}%
\right) =\left( 
\begin{array}{cccc}
0 & 0 & 1 & 0 \\ 
0 & 0 & 0 & 1 \\ 
-V_{11} & -V_{12} & 0 & 0 \\ 
-V_{12} & -V_{22} & 0 & 0%
\end{array}%
\right) \left( 
\begin{array}{c}
\delta x_{1} \\ 
\delta x_{2} \\ 
\delta y_{1} \\ 
\delta y_{2}%
\end{array}%
\right)  \tag{A3}
\end{equation}

where $V_{ij}\equiv \frac{\partial ^{2}V}{\partial x_{i}\partial x_{j}}.$
The eigenvalues (here $\lambda $ is the standard Lyapunov eigenvalue) are
given by the roots of

\begin{equation}
\lambda ^{4}+\lambda ^{2}\left( V_{11}+V_{22}\right) +\left(
V_{11}V_{22}-V_{12}^{2}\right) =0  \tag{A4}
\end{equation}

Compare these eigenvalues to those of the Hamiltonian stability matrix $%
\mathcal{V}^{HS}=\left( V_{ij}\right) ,$ Eq.(21),

\begin{equation}
\left\vert 
\begin{array}{cc}
V_{11}-\lambda _{HS} & V_{12} \\ 
V_{12} & V_{22}-\lambda _{HS}%
\end{array}%
\right\vert =0  \tag{A5}
\end{equation}

Hence%
\begin{equation}
\lambda _{HS}^{2}-\lambda _{HS}\left( V_{11}+V_{22}\right) +\left(
V_{11}V_{22}-V_{12}^{2}\right) =0  \tag{A6}
\end{equation}

With $\lambda _{HS}$ replaced by $-\lambda ^{2}$ Eq.(A6) becomes equivalent
to Eq.(A4), so that $\lambda _{HS}=-\lambda ^{2}$.

The stability conditions are 
\begin{equation}
\begin{array}{c}
\lambda _{HS}>0,\ \ \ \ \ \lambda ^{2}<0\ \ \ \rightarrow \ \ \ \ \ \text{%
stable} \\ 
\lambda _{HS}<0,\ \ \ \ \ \lambda ^{2}>0\ \ \ \rightarrow \ \ \ \ \ \text{%
unstable}%
\end{array}
\tag{A7}
\end{equation}

Hence in our notation, positive eigenvalues of $\frac{\partial ^{2}V}{%
\partial x_{i}\partial x_{j}}$ result in Lyapunov stability, negative
eigenvalues of $\frac{\partial ^{2}V}{\partial x_{i}\partial x_{j}}$ result
in Lyapunov instability.


\begin{thebibliography}{99}
\bibitem{Berry} We wish to thank Michael Berry for bringing this point to
our attention.
\end{thebibliography}
\end{document}